\documentclass[twocolumn,english,aps,prb,twocolum,superscriptaddress,bibnotes,amsmath,amssymb,floatfix]{revtex4-1}
\usepackage[colorlinks=true,citecolor=blue,linkcolor=magenta]{hyperref}

\usepackage[markup=nocolor, authormarkupposition=left]{changes} 
\usepackage{soul}
\usepackage[utf8]{inputenc}
\usepackage[english]{babel}
\usepackage{amsmath,amsfonts,amssymb}
\usepackage[T1]{fontenc}
\usepackage{url}

\usepackage{amsmath}
\usepackage{amsfonts}
\usepackage{amssymb}

\usepackage{epstopdf}
\usepackage{graphicx}
\graphicspath{{./Figures/}}

\usepackage{changes}

\begin{document}
\title{A chip-based optoelectronic-oscillator frequency comb}

\author{Jinbao Long}
\thanks{These authors contributed equally to this work.}
\affiliation{International Quantum Academy, Shenzhen 518048, China}

\author{Zhongkai Wang}
\thanks{These authors contributed equally to this work.}
\affiliation{International Quantum Academy, Shenzhen 518048, China}

\author{Huanfa Peng}
\thanks{These authors contributed equally to this work.}
\affiliation{Institute of Photonics and Quantum Electronics (IPQ), Karlsruhe Institute of Technology (KIT), Karlsruhe 76131, Germany}

\author{Wei Sun}
\affiliation{International Quantum Academy, Shenzhen 518048, China}

\author{Dengke Chen}
\affiliation{International Quantum Academy, Shenzhen 518048, China}
\affiliation{Shenzhen Institute for Quantum Science and Engineering, Southern University of Science and Technology, Shenzhen 518055, China}

\author{Shichang Li}
\affiliation{International Quantum Academy, Shenzhen 518048, China}
\affiliation{Shenzhen Institute for Quantum Science and Engineering, Southern University of Science and Technology, Shenzhen 518055, China}

\author{Shuyi Li}
\affiliation{International Quantum Academy, Shenzhen 518048, China}

\author{Yi-Han Luo}
\affiliation{International Quantum Academy, Shenzhen 518048, China}

\author{Lan Gao}
\affiliation{International Quantum Academy, Shenzhen 518048, China}

\author{Baoqi Shi}
\affiliation{International Quantum Academy, Shenzhen 518048, China}

\author{Chen Shen}
\affiliation{International Quantum Academy, Shenzhen 518048, China}
\affiliation{Qaleido Photonics, Shenzhen 518048, China}

\author{Jijun He}
\affiliation{Key Laboratory of Radar Imaging and Microwave Photonics, Ministry of Education, Nanjing University of Aeronautics and Astronautics, Nanjing 210016, China}

\author{Linze Li}
\affiliation{School of Information Science and Technology, ShanghaiTech University, Shanghai 201210, China}

\author{Tianyu Long}
\affiliation{School of Information Science and Technology, ShanghaiTech University, Shanghai 201210, China}

\author{Baile Chen}
\affiliation{School of Information Science and Technology, ShanghaiTech University, Shanghai 201210, China}

\author{Zhenyu Li}
\affiliation{Institute of Microelectronics, Agency for Science, Technology and Research (A*STAR), Singapore}

\author{Junqiu Liu}
\email[]{liujq@iqasz.cn}
\affiliation{International Quantum Academy, Shenzhen 518048, China}
\affiliation{Hefei National Laboratory, University of Science and Technology of China, Hefei 230088, China}

\maketitle

\noindent\textbf{Microresonator-based Kerr frequency combs (``Kerr microcombs'') constitute chip-scale frequency combs of broad spectral bandwidth and repetition rate ranging from gigahertz to terahertz. 
An appealing application exploiting microcombs’ coherence
and large repetition rate is microwave and millimeter-wave generation. 
Latest endeavor applying two-point optical frequency division (OFD) on photonic-chip-based microcombs has created microwaves with exceptionally low phase noise.
Nevertheless, microcomb-based OFD still requires extensive active locking, additional lasers, and external RF or microwave sources, as well as sophisticated initiation. 
Here we demonstrate a simple and entirely passive (no active locking) architecture, which incorporates an optoelectronic oscillator (OEO) and symphonizes a coherent microcomb and a low-noise microwave spontaneously.  
Our OEO microcomb leverages state-of-the-art integrated chip devices including a high-power DFB laser, a broadband silicon Mach-Zehnder modulator, an ultralow-loss silicon nitride microresonator, and a high-speed photodetector. 
Each can be manufactured in large volume with low cost and high yield using established CMOS and III-V foundries. 
Our system synergizes a microcomb of 10.7 GHz repetition rate and an X-band microwave with phase noise of $-$97/$-$126/$-$130 dBc/Hz at 1/10/100 kHz Fourier frequency offset, yet does not demand active locking, additional lasers, and external RF or microwave sources. 
With potential to be fully integrated, our OEO microcomb can become an invaluable technology and building block for microwave photonics, radio-over-fiber, and optical communication.}

\begin{figure*}[t!]
\centering
\includegraphics{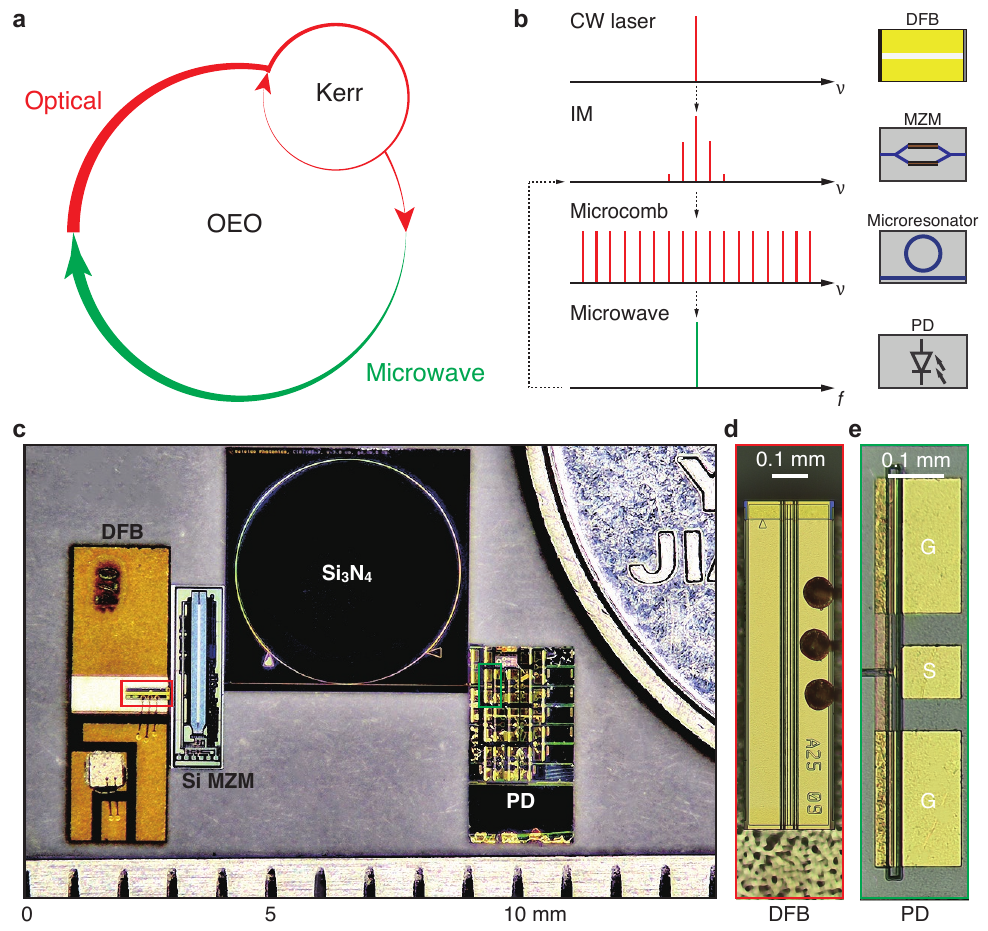}
\caption{
\textbf{Concept, principle and components of the optoelectronic-oscillator microcomb.}
\textbf{a}. 
Conceptual diagram of the self-starting and self-maintaining OEO microcomb via the feedback interplay of Kerr nonlinearity and optical-microwave conversion.
\textbf{b}. 
Principle of the OEO microcomb.
The DFB laser's CW output is intensity-modulated (IM) by an MZM. 
The CW pump and modulated sidebands are coupled into a high-$Q$ optical microresonator. 
If the IM frequency $f_\mathrm{IM}$ matches the microresonator FSR, i.e. $f_\mathrm{IM}=D_1/2\pi$, a coherent microcomb forms, whose line spacing is $f_\mathrm{rep}=D_1/2\pi$. 
Detection of the microcomb's $f_\mathrm{rep}$ via a PD outputs a microwave of carrier frequency $f_\mathrm{rep}$, which is injected to the MZM. 
As such, $f_\mathrm{IM}=f_\mathrm{rep}=D_1/2\pi$ is ensured. 
Consequently, the entire system can self-oscillate and self-maintain, harmonizing a coherent microcomb and a low-noise microwave. 
\textbf{c}. 
Photograph of the DFB laser chip, the Si MZM chip, the Si$_3$N$_4$ microresonator chip, and the PD chip, referenced to a ruler and in comparison with the size of a 1-Chinese-Jiao coin. 
\textbf{d}.
Optical microscope image of the DFB laser. 
\textbf{e}.
Optical microscope image of the PD with GSG pads. 
}
\label{Fig:1}
\end{figure*}

\begin{figure*}[t!]
\centering
\includegraphics{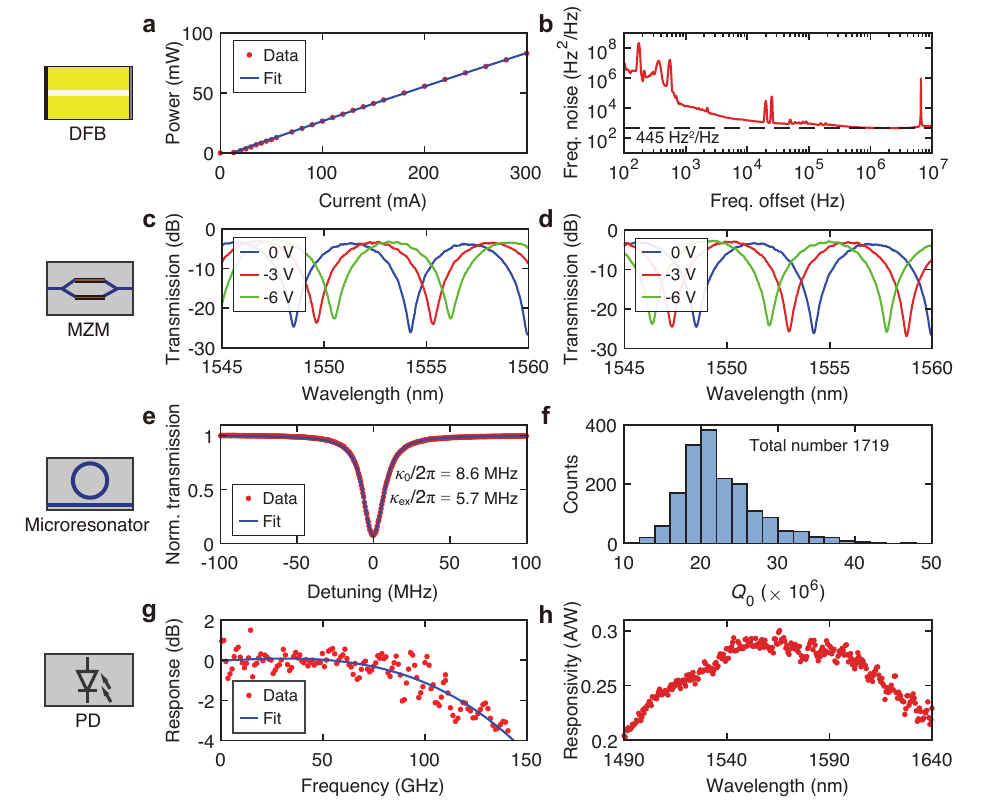}
\caption{
\textbf{Characterization of chips.}
\textbf{a, b}. 
Measured output optical power versus laser current (a), 
and single-sideband  frequency noise PSD (b) of the free-running DFB laser. 
The 445 Hz$^2/$Hz white noise corresponds to 2.79 kHz intrinsic linewidth. 
\textbf{c, d}.
Measured transmission spectra of the Si TW-MZM with different bias voltage $V_\text{dc}$ applied on the upper (c) or lower waveguide arm (d).
\textbf{e, f}. 
A typical resonance of the Si$_3$N$_4$ microresonator with fitted $\kappa_0/2\pi=8.6$ MHz and $\kappa_\text{ex}/2\pi=5.7$ MHz (e), 
and the histogram of 1,719 measured $Q_0$ values with the most probable value $Q_0=21\times10^6$ (f). 
\textbf{g, h}. 
Measured detection bandwidth (g) and responsivity versus wavelength (h) of the PD chip. 
The 3-dB bandwidth is estimated over 120 GHz from the fitting.
}
\label{Fig:2}
\end{figure*}

Optical frequency combs (OFC) \cite{Udem:02, Cundiff:03, Fortier:19}, which coherently channel radio- and microwave frequency to optical domain, have revolutionized timing, spectroscopy, and precision measurement, as well as test of fundamental physics. 
Conventionally constructed with solid-state or fiber mode-locked lasers, today OFCs can be built on-chip \cite{Kippenberg:18, Gaeta:19, Diddams:20}. 
Such remarkable advancement has been made possible by the emergence and quick maturing of low-loss photonic integrated circuit based on a variety of material platforms \cite{Moss:13, Kovach:20, Dutt:24}, along with hybrid and heterogeneous integration \cite{Komljenovic:16, Stern:18, Xiang:21}. 
Photonic-chip-based OFCs feature small size, weight and power consumption, and can be manufactured in large volume with low cost and high yield, ideal for wide deployment outside laboratories and in space. 

The most leading type of photonic-chip-based OFCs is established on low-loss, Kerr-nonlinear optical microresonators driven by continuous-wave (CW) lasers, which is commonly referred to as ``Kerr microcombs'' \cite{DelHaye:07, Herr:14, Brasch:15, Joshi:16, Liang:15, Yi:15, Liu:20, ZhangS:19a, Tetsumoto:21, Xue:15, Lobanov:15, Huang:15b, Parra-Rivas:16, Nazemosadat:21}. 
Microcombs exhibit broad spectral bandwidth and repetition rates in the gigahertz to terahertz range. 
One critical application benefiting from the coherence and large repetition rate of microcombs is microwave and millimeter-wave generation \cite{Liang:15, Yi:15, Liu:20, ZhangS:19a, Tetsumoto:21}. 
Photodetection of the microcomb pulse stream generates a low-noise microwave or millimeter-wave whose carrier frequency corresponds to the microcomb's repetition rate. 
Various approaches have been demonstrated to improve the microwave's spectral purity, aided by an external microwave \cite{Weng:19}, an auxiliary laser \cite{LiuR:24}, a transfer comb \cite{Lucas:20}, or operation in the ``quiet point'' \cite{Yi:17, Yang:21}. 
Notably, latest endeavor has applied optical frequency division (OFD) \cite{Fortier:11, Xie:17, Li:14} on photonic-chip-based microcombs, catalyzing microwaves with superior phase noise performance \cite{Kudelin:24, SunS:24, Zhao:24, He:24, Jin:24}.
Nevertheless, microcomb-based OFD still requires extensive active locking, additional lasers, and external RF or microwave sources, as well as sophisticated initiation. 

Here we demonstrate an architecture combining an optoelectronic oscillator (OEO) and a microcomb. 
Figure \ref{Fig:1}a depicts the conceptual diagram of our ``OEO microcomb''. 
An OEO comprises hybrid optical and microwave components, forming a photonic microwave oscillator with ultra-high RF spectral purity \cite{Maleki:11, Tang:18, Peng:21, Chembo:19, Hao:20a}. 
By embedding a high-$Q$ microresonator in the optical part of the OEO loop, the feedback interplay of the Kerr nonlinearity and the self-sustained microwave oscillation symphonizes a microcomb and a microwave spontaneously. 

Figure \ref{Fig:1}b illustrates the working principle of our OEO microcomb. 
The CW output from a distributed feedback (DFB) laser is intensity-modulated with a silicon Mach-Zehnder modulator (Si MZM), creating pairs of optical sidebands in the frequency domain.  
The modulated light is coupled into a high-$Q$ silicon nitride (Si$_3$N$_4$) optical microresonator. 
When the MZM's modulation frequency $f_\mathrm{IM}$ matches the microresonator's free spectral range (FSR, $D_1/2\pi$), i.e. $f_\mathrm{IM}=D_1/2\pi$, a coherent microcomb forms \cite{Cole:18, Miao:22, Lobanov:19, Liu:22}. 
In the frequency domain, the microcomb contains many mutually coherent CW tones that are equidistantly spaced by $D_1/2\pi$. 
In the time domain, it is a pulse stream of repetition rate $f_\mathrm{rep}=D_1/2\pi$. 
Detection of the microcomb via a photodetector (PD) outputs a fundamental microwave tone with carrier frequency of $f_\mathrm{rep}$. 
Collecting and injecting the microwave back to the Si MZM ensures $f_\mathrm{IM}=f_\mathrm{rep}=D_1/2\pi$. 
As such, the entire system can self-oscillate and self-maintain, harmonizing a coherent microcomb and a low-noise microwave. 
Figure \ref{Fig:1}c presents a photograph of the four chip components of our OEO microcomb -- a DFB laser chip, a Si MZM chip, a Si$_3$N$_4$ microresonator chip, and a PD chip.
The chip sizes are referenced to a ruler. 
Figure \ref{Fig:1}d and e show the zoomed-in optical microscope images of the III-V-semiconductor-based DFB laser chip and an individual PD on the PD chip. 
Details on the performance characterization of each chip components are summarized in Fig. \ref{Fig:2} and described in the following.  

\section*{Characterization of individual chip components}
\noindent\textbf{Laser}. 
The commercial DFB laser outputs 83 mW CW power around 1558 nm with 300 mA current and transverse-electric (TE) polarization. 
A printed circuit board (PCB) is used to stabilize the laser's current and temperature at 25$^\circ$C. 
Figure~\ref{Fig:2}a shows that the laser exhibits 13 mA current at threshold, and 0.6 nm wavelength tunability over 300 mA current range.
Figure~\ref{Fig:2}b shows the measured frequency noise of the free-running laser, i.e. single-sideband power spectrum density (PSD) of the laser frequency noise.
The intrinsic linewidth is calculated as 2.79 kHz from the white noise of 445 Hz$^2/$Hz.
More information of the DFB laser is found in Supplementary Materials Note 1.

\noindent\textbf{Modulator}.
The Si MZM, fabricated in a standard CMOS foundry, is a travelling-wave (TW) MZM with push-pull configuration \cite{Li:25}. 
Modulation is achieved via plasma dispersion effect in depletion-type PN junctions within the two waveguide arms. 
The TW-MZM's microwave-to-optic conversion efficiency is characterized by the product of half-wave voltage $V_\pi$ and phase-shift length $L_\pi$, i.e. $V_\pi L_\pi$, which can be measured by applying a bias voltage $V_\text{dc}$ on the upper or lower waveguide arm of the TW-MZM. 
Figure~\ref{Fig:2}c and d show the Si TW-MZM's transmission spectrum with different $V_\text{dc}$ values applied on the upper (c) or lower arm (d).
By calculating the dip shift on the transmission spectrum under varying $V_\text{dc}$, the $V_\pi L_\pi$ of our Si TW-MZM is calculated as 2 V$\cdot$cm. 
More information of the Si TW-MZM is found in Supplementary Materials Note 2. 

\noindent\textbf{Microresonator}.
The Si$_3$N$_4$ microresonator is fabricated using a foundry-level, deep-ultraviolet subtractive process with 300-nm-thick Si$_3$N$_4$ on 150-mm-diameter (6-inch) wafers \cite{Ye:23, Sun:24}. 
Light is coupled into and out of the Si$_3$N$_4$ microresonator's fundamental TE mode via inverse tapers at chip facets and a bus waveguide.
At 1558 nm pump wavelength, the Si$_3$N$_4$ microresonator features $D_1/2\pi=10.699$ GHz FSR. 
The microresonator's $Q$ factor is evaluated by resonance fit \cite{Luo:24}. 
Figure~\ref{Fig:2}e presents a typical resonance with fitted intrinsic loss $\kappa_0/2\pi=8.6$ MHz, external coupling strength $\kappa_\text{ex}/2\pi=5.7$ MHz, and loaded linewidth $\kappa/2\pi=(\kappa_0+\kappa_\text{ex})/2\pi=14.3$ MHz. 
The intrinsic quality factor is calculated as  $Q_0=\omega/\kappa_0$, where $\omega/2\pi$ is the resonant frequency.
Figure~\ref{Fig:2}f shows the histogram of 1,719 measured $Q_0$ values, with the most probable value $Q_0=21\times10^6$. 
As the 300 nm Si$_3$N$_4$ thickness endows the microresonator with normal group velocity dispersion (GVD, $D_2<0$), the generated microcomb is a dark pulse (platicon) stream in the time domain \cite{Xue:15, Lobanov:15, Huang:15b, Parra-Rivas:16, Nazemosadat:21}. 
The fabrication process flow and more characterization data of the Si$_3$N$_4$ microresonator are found in Supplementary Materials Note 3. 

\noindent\textbf{Photodetector}. 
The PD chip has $3\times15$ $\mu$m$^2$ active area, whose epitaxial structure is grown on a semi-insulating indium phosphide (InP) substrate \cite{LiL:23}.
It collects incident light via a waveguide, and outputs electrical signals via a ground-signal-ground (GSG) probe.
Figure~\ref{Fig:2}g shows the measured 3-dB bandwidth over 120 GHz. 
Figure~\ref{Fig:2}h shows that, within the detection wavelength range from 1490 to 1640 nm, the measured responsivity is above 0.2 A/W and up to 0.3 A/W. 
The fabrication process flow and more characterization data of the PD chip are found in Supplementary Materials Note 4.

\begin{figure*}[t!]
\centering
\includegraphics{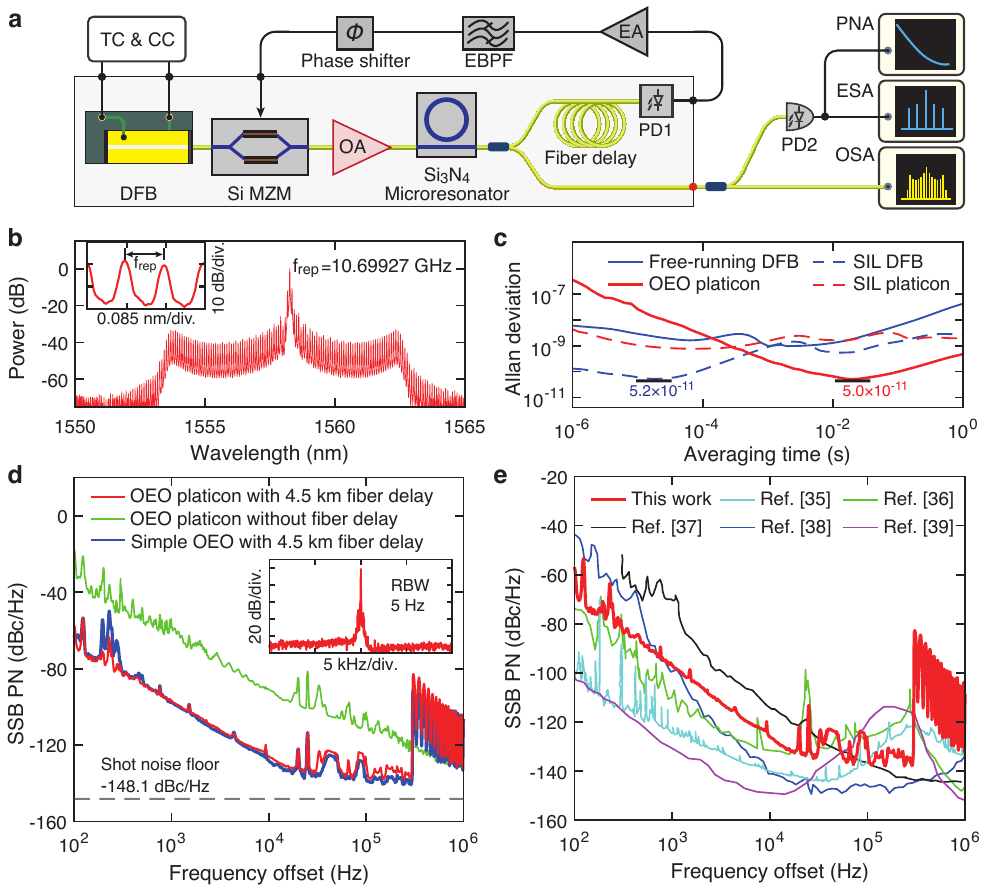}
\caption{
\textbf{Experimental setup and results of the optoelectronic-oscillator microcomb.}
\textbf{a}. 
Experimental setup. 
TC $\&$ CC, temperature control and current control.  
ESA, electrical spectrum analyser. 
OSA, optical spectrum analyser.
PD1, the PD chip.
PD2, the Finisar PD.
\textbf{b}. 
Optical spectrum of the platicon microcomb. 
Inset shows 10.69927 GHz line spacing.
\textbf{c}. 
Allan deviation data of the free-running DFB laser's frequency (optical, blue solid curve), and the OEO platicon's $f_\mathrm{rep}$ (microwave, red solid curve), in comparison with the SIL DFB laser's frequency (optical, blue dashed curve), and the SIL palticon's $f_\mathrm{rep}$ (microwave, red dashed curve) described in Ref. \cite{Sun:24}. 
\textbf{d}.
Phase noise data of the $f_\mathrm{rep}=10.69927$ GHz microwaves of the OEO platicon with (red curve) and without (green curve) 4.5 km fiber delay, in comparison with the conventional OEO with 4.5 km fiber delay (blue curve) and the estimated shot noise floor (dashed black curve). 
Inset shows the RF spectrum of the platicon microcomb's $f_\mathrm{rep}$. 
RBW, resolution bandwidth. 
\textbf{e}.
Phase noise data of our OEO platicon microcomb's $f_\mathrm{rep}=10.69927$ GHz (red curve), 
in comparison with the phase noise data using microcomb-based OFD (scaled to 10 GHz) in Ref.\cite{Kudelin:24} (cyan curve), 
Ref.\cite{SunS:24} (green curve), 
Ref.\cite{Zhao:24} (black curve), 
Ref.\cite{He:24} (blue curve),
Ref.\cite{Jin:24} (magenta curve). 
}
\label{Fig:3}
\end{figure*}

\section*{Experimental results}
\noindent\textbf{Experimental setup}.
Figure \ref{Fig:3}a shows the experimental setup including optical and electronic components that are not integrated, besides the four types of chips described above. 
The DFB laser's CW output at 1558 nm wavelength is sent into the Si MZM.
The intensity-modulated light from the Si MZM is power-boosted via an optical amplifier (OA), which in our case is an EDFA. 
The amplified light is then coupled into the Si$_3$N$_4$ microresonator and generates a platicon microcomb. 
Before detection with the high-speed PD chip (PD1), the platicon pulse stream travels through a 4.5-km-long single-mode-fiber (SMF) whose function will be discussed later. 
The PD chip outputs a microwave signal whose carrier frequency corresponds to the platicon's repetition rate $f_\mathrm{rep}$. 
The microwave signal is power-boosted by an electronic amplifier (EA) and filtered by an electronic band-pass filter (EBPF). 
The EBPF's center frequency is fixed at $f_\mathrm{rep}=D_1/2\pi=10.699$ GHz and 3-dB bandwidth is 10 MHz. 
A phase shifter is used to vary the feedback microwave signal's phase before injecting the signal to drive the Si MZM. 

With optimized combination of the DFB laser frequency, the EA gain, and the feedback microwave phase, a platicon microcomb and a low-noise microwave can synergetically self-start from the noise in the OEO loop, and self-maintain. 
This is due to that the OEO satisfies the classical Barkhausen oscillation criteria of feedback-loop systems \cite{Chembo:19}. 
Because of the 4.5-km-long fiber delay, the OEO function endows the microwave with high spectral purity (i.e. low phase noise), which further purifies the modulation signal for the CW pump and improves the platicon's coherence.

Experimentally, the EA contains a two-stage low-noise amplifier (maximum 46 dB gain) and a voltage-variable attenuator (VVA, maximum 50 dB attenuation). 
The phase shifter is voltage-controlled, allowing for 0$^\circ$ to 360$^\circ$ loop phase variation. 
We treat 0$^\circ$ relative phase with 0 V applied.
When the EA gain is set as 36 dB and the loop phase is 240$^\circ$, the system self-oscillates.
The platicon microcomb's optical spectrum is shown in Fig.~\ref{Fig:3}b. 
The inset highlights comb lines with 10.69927 GHz spacing.
The ultimate microwave power applied on the Si MZM is measured as 19 dBm. 
We observe that the platicon microcomb can self-maintain with microwave power ranging from 16 to 22 dBm by varying the EA gain. 

Previously, platicon microcombs in optical microresonators can be deterministically seeded by phase- or intensity-modulation of the CW pump \cite{Lobanov:19, Liu:22}.  
These methods necessitate an external microwave source operating at a high frequency to match the microresonator FSR, which is bulky, power-hungry and expensive. 
Here we obviate external microwave sources by utilizing the microwave synthesized via photodetection of the platcon's $f_\mathrm{rep}$ and the OEO loop. 
In addition, the microwave frequency is selected by the EBPF's center frequency -- a passive component that is simple, small, stable and does not consume power.

\noindent\textbf{Coherence optimization.} 
The mutual coherence of the platicon's comb lines is characterized by photodetection of the platicon's $f_\mathrm{rep}$ using another PD (PD2, Finisar XPDV3120R-VF-FA) and analysis of single-sideband RF phase noise (SSB PN) of $f_\mathrm{rep}$ using a phase noise analyzer (PNA).
Figure \ref{Fig:3}d green curve shows the measured microwave phase noise of the OEO platicon's $f_\mathrm{rep}$ without fiber delay, which reaches $-57$/$-91$/$-104$ dBc/Hz at 1/10/100 kHz Fourier frequency offset.
To purify the microwave, a 4.5-km-long fiber delay is introduced, as exemplified in Ref. \cite{Eliyahu:08}.  
Typically, a longer fiber leads to lower microwave phase noise. 
However there are parasitic limitations on the allowed maximum fiber length.
First, a longer fiber results in smaller OEO mode spacing. 
In our case with 4.5 km fiber length, the mode spacing is 43.6 kHz. 
Second, the side-mode suppression ratio (SMSR) decreases as the fiber length increases.  
For example, the SMSR is around 35(50) dB for 4.5(2.0) km fiber length.  
These two effects together cause multi-mode competition, resulting in instability, mode-hopping and extra noise. 
The EBPF of 10 MHz narrow bandwidth ameliorates this issue and allows stable oscillation with 4.5 km fiber length. 
Measured OEO microwave's phase noise with different fiber delay length is compared in Supplementary Materials Note 5. 

Finally, with 4.5 km fiber delay, the phase noise of our OEO platicon's $f_\mathrm{rep}$ reaches $-97$/ $-126$/$-130$ dBc/Hz at 1/10/100 kHz Fourier frequency offset, as shown in Fig.~\ref{Fig:3}d red curve. 
The inset shows the electrical power spectrum of the $f_\mathrm{rep}=10.69927$ GHz microwave with 5 Hz resolution bandwidth.
In comparison, we also measure the conventional OEO microwave's phase noise without platicon formation and with 4.5 km fiber delay, as shown in Fig.~\ref{Fig:3}d blue curve. 
The microwave phase noises with and without platicon formation are nearly identical, indicating that $f_\mathrm{rep}$ directly inherits coherence  from the OEO microwave. 
Thus we do not observe platicon-induced spectral purification effect described in Ref. \cite{Liu:22}. 
The reason is due to that our microwave phase noise is already sufficiently low. 
More experimental data to quantify the platicon-induced spectral purification effect is found in Supplementary Materials Note 6.

Figure~\ref{Fig:3}c shows the measured Allan deviation data of the free-running DFB laser's frequency (optical, blue solid curve), 
and the OEO platicon's $f_\mathrm{rep}$ (microwave, red solid curve), using the PNA. 
In comparison, the Allan deviation data of the self-injection-locked (SIL) DFB laser's frequency (optical, blue dashed curve), 
and the SIL palticon's $f_\mathrm{rep}$ (microwave, red dashed curve), from Ref. \cite{Sun:24}, are shown. 
It is evident that the frequency instability of the OEO platicon's $f_\mathrm{rep}$ is significantly lower than that of the SIL platicon's $f_\mathrm{rep}$, though the free-running DFB laser is less stable than the SIL DFB laser.
The long-term drift of our OEO platicon's $f_\mathrm{rep}$ is mainly determined by the DFB laser's frequency drift and the thermal drift of the long fiber delay. 

The phase noise of $f_\mathrm{rep}$ increases from $-126$ dBc/Hz to $-118$ dBc/Hz at 10 kHz offset by varying the EA gain.  
A large gain can introduce bumps on the phase noise curve as detailed in Supplementary Materials Note 7. 
Reference \cite{Chembo:19} illustrates the loop gain's influence on the bifurcation sequence in the OEO. 
There is an optimum loop-gain range such that Hopf bifurcation with constant amplitude generates an ultra-pure microwave. 
Such an optimum gain is also found in our experiment.
Meanwhile, Supplementary Materials Note 8 investigates possible reasons that limit our microwave phase noise. 
For example, the phase noise can be further reduced to $-102$/$-130$/$-131$ dBc/Hz at 1/10/100 kHz offset by replacing the chip PD with a commercial PD (another Finisar XPDV3120R-VF-FA). 
Meanwhile, replacing the Si MZM with a commercial lithium niobate electro-optic modulator (LiNbO$_3$ EOM, iXblue MXAN-LN-10-PD) reduces the phase noise at low Fourier frequency offset, e.g. 0.1 kHz.
The reasons are likely due to that both the Si MZM and the chip PD are not fully packaged, where optical and microwave power fluctuation on these devices causes loop gain jittering. 
In addition, we do not observe phase noise reduction by replacing the DFB laser with an external-cavity diode laser (ECDL, Toptica CTL). 

Figure \ref{Fig:3}e compares our 10.69927 GHz microwave's phase noise with scaled 10 GHz microwave's phase noise from recent microcomb-based OFD works \cite{Kudelin:24, SunS:24, Zhao:24, He:24, Jin:24}. 
Different from Ref. \cite{Kudelin:24, SunS:24, Zhao:24, He:24, Jin:24}, our OEO-platicon-based microwave generation does not require any active locking, servos, multiple lasers, and external RF or microwave sources for reference or regulation. 
\begin{figure*}[t!]
\centering
\includegraphics{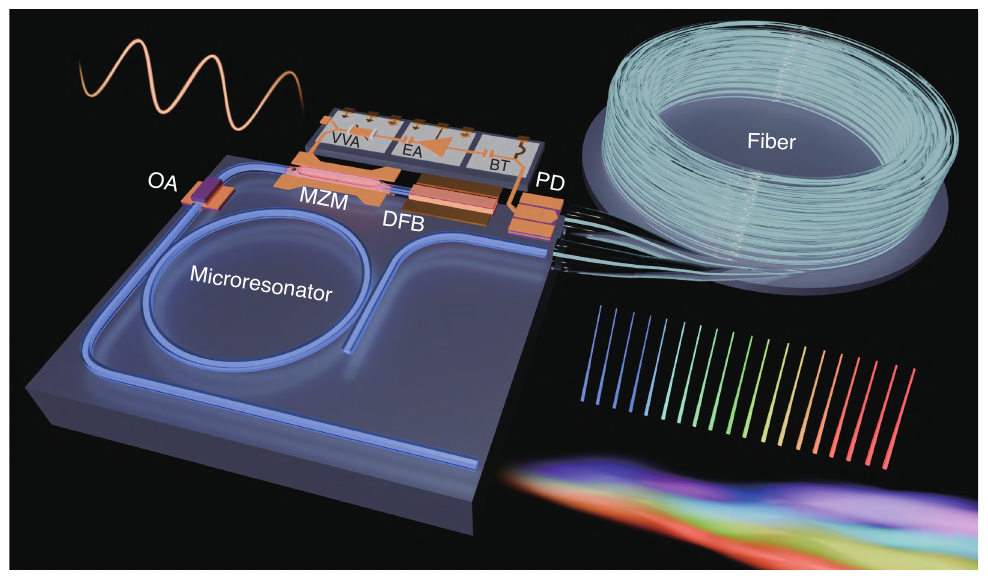}
\caption{
\textbf{Envisaged architecture of a fully integrated OEO microcomb.}
BT, bias-tee. 
}
\label{Fig:4}
\end{figure*}

\section*{Discussion and conclusion}

\noindent\textbf{Towards heterogeneous integration.} 
While the DFB laser, the Si MZM, the Si$_3$N$_4$ microresonator, and the PD are photonic chips, the entire OEO platicon system is not yet fully integrated. 
Nevertheless, with maturing heterogeneous integration and photonic-electronic co-packaging, our OEO microcomb system has the potential to be fully integrated \cite{Xiang:24}. 
Figure \ref{Fig:4} depicts an envisaged, fully integrated architecture. 
The photonic Damascene process \cite{Liu:21} is advantageous to fabricate ultralow-loss Si$_3$N$_4$ waveguides and microresonators with planarized top surface, ideal for wafer bonding of other thin films \cite{Chang:17, Xiang:21} and micro-transfer printing \cite{Roelkens:24}. 
On Si$_3$N$_4$ waveguides, DFB lasers and PDs can be heterogeneously integrated, as described in Ref. \cite{Xiang:21} and Ref. \cite{Yu:20}, respectively. 
In addition to Si MZMs which are naturally compatible with heterogeneous integration on Si$_3$N$_4$, thin-film LiNbO$_3$ EOMs can also be integrated on Si$_3$N$_4$ as demonstrated in Ref. \cite{Churaev:23}. 
The optical amplifier can be erbium-doped Si$_3$N$_4$ waveguides \cite{LiuY:22} or III-V semiconductor optical amplifiers \cite{OpdeBeeck:20}. 

Meanwhile, ultralow-loss waveguides and microresonators can also be made of LiNbO$_3$ \cite{Zhu:21, Zhang:17}, allowing microwave-rate bright soliton \cite{He:23} or dark pulse \cite{Lv:24} generation.  
In such, the inclusion of EOMs is straight-forward \cite{WangC:18, Ma:24} and does not require heterogeneous integration. 
Lasers and PDs can be heterogeneously integrated on LiNbO$_3$ \cite{Li:24}. 
The optical amplifier can be erbium-doped LiNbO$_3$ waveguides \cite{ChenZ:21} or III-V semiconductor optical amplifiers \cite{OpdeBeeck:21}. 

As it is impossible to fabricate kilometer-long integrated waveguides, fiber can only be hybrid-integrated or packaged to the OEO microcomb chip, as well as the microelectronic chip containing the VVA, EA and bias-tee. 
In our experiment, the tightly winded fiber has a volume below 500 cm$^3$. 
In addition, Fig. \ref{Fig:4} shows that the microresonator with a drop port can serve as an optical band-pass filter in the OEO. 
In such, the EBPF can be excluded and the required fiber length can be significantly shortened with the high-$Q$ microresonator \cite{Ma:24}. 
This leads to not only reduced size and weight, but also a more stable microcomb with relaxed OEO multi-mode competition. 
Besides, the drop port can also suppress amplified spontaneous emission (ASE) noise.

\noindent\textbf{Conclusion.} 
In conclusion, we have demonstrated an OEO microcomb which spontaneously harmonizes a coherent microcomb and a low-noise microwave. 
Critical components of our OEO microcomb involve a high-power DFB laser, a broadband Si MZM, an ultralow-loss Si$_3$N$_4$ microresonator, and a high-speed PD. 
Each component represents the state of the art in its own class, yet can be manufactured in large volume with low cost and high yield using established CMOS and III-V foundries. 
The synthesized microcomb features 10.7 GHz repetition rate with phase noise of $-97$/$-126$/$-130$ dBc/Hz at 1/10/100 kHz Fourier frequency offset. 
In contrast to recent demonstrations using microcomb-based OFD \cite{Kudelin:24, SunS:24, Zhao:24, He:24, Jin:24}, our OEO microcomb can achieve comparable phase-noise performance \cite{Zhao:24, He:24}, yet is entirely passive and much simpler. 
Thus our work paves a greatly simplified route to symphonizing coherent and robust microcombs and microwaves.
Moreover, with the fast evolving heterogeneous and hybrid integration, our OEO microcomb has promising potential to be fully integrated on a monolithic chip with improved stability. 
Our OEO microcomb can become an invaluable technology and building block for microwave photonics, radio-over-fiber, and optical communication.

\medskip
\begin{footnotesize}
\noindent \textbf{Funding Information}: 
We acknowledge support from the National Natural Science Foundation of China (Grant No. 12261131503, 12404436, 12404417, 62405202, 61975121, 62205145), 
Innovation Program for Quantum Science and Technology (2023ZD0301500), 
National Key R\&D Program of China (Grant No. 2024YFA1409300), 
Shenzhen Science and Technology Program (Grant No. RCJC20231211090042078), 
Shenzhen-Hong Kong Cooperation Zone for Technology and Innovation (HZQB-KCZYB2020050), 
and Guangdong-Hong Kong Technology Cooperation Funding Scheme (Grant No. 2024A0505040008). 

\vspace{0.1cm}
\noindent \textbf{Acknowledgments}: 
We thank Zhiyang Chen for assistance in the experiment, and Fuchuan Lei for inspiring discussion. 
L.L, T.L. and B. C. are grateful to the device fabrication support from the ShanghaiTech University Material Device Lab (SMDL).
Silicon nitride chips were fabricated by Qaleido Photonics. 
The PD chips were fabricated with support from the ShanghaiTech University Quantum Device Lab (SQDL).

\vspace{0.1cm}
\noindent \textbf{Author contributions}: 
J. Long, Z. W., W. S. and J. H. built the experimental setup. 
Z. W., J. Long, H. P. and W. S. performed the experiments and analyzed the data, with the assistance from D. C., Shichang L., Y.-H. L. and J. H.. 
L. G., B. S. and C. S. fabricated and characterized the Si$_3$N$_4$ chip. 
Shuyi L. and Z. L. fabricated and characterized the Si MZM chip.
L. L, T. L. and B. C. fabricated and characterized the PD chip. 
J. Long, W. S., and Shichang L. characterized and packaged the DFB lasers. 
Z. W., W. S., J. Long, H. P. and J. Liu wrote the manuscript, with the input from others. 
J. Liu initiated the collaboration and supervised the project.

\vspace{0.1cm}
\noindent \textbf{Disclosures}: 
J. Long, Z. W., W. S. and J. Liu are inventors on a patent application related to this work. 
C. S. and J. Liu are co-founders of Qaleido Photonics, a start-up that is developing heterogeneous silicon nitride integrated photonics technologies. 
Others declare no conflicts of interest.

\vspace{0.1cm}
\noindent \textbf{Data Availability Statement}: 
The code and data used to produce the plots within this work will be released on the repository \texttt{Zenodo} upon publication of this preprint.

\end{footnotesize}

\end{document}


\title{Supplementary Materials for:
A chip-based optoelectronic-oscillator frequency comb}

\author{Jinbao Long}
\thanks{These authors contributed equally to this work.}
\affiliation{International Quantum Academy, Shenzhen 518048, China}

\author{Zhongkai Wang}
\thanks{These authors contributed equally to this work.}
\affiliation{International Quantum Academy, Shenzhen 518048, China}

\author{Huanfa Peng}
\thanks{These authors contributed equally to this work.}
\affiliation{Institute of Photonics and Quantum Electronics (IPQ), Karlsruhe Institute of Technology (KIT), Karlsruhe 76131, Germany}

\author{Wei Sun}
\affiliation{International Quantum Academy, Shenzhen 518048, China}

\author{Dengke Chen}
\affiliation{International Quantum Academy, Shenzhen 518048, China}
\affiliation{Shenzhen Institute for Quantum Science and Engineering, Southern University of Science and Technology, Shenzhen 518055, China}

\author{Shichang Li}
\affiliation{International Quantum Academy, Shenzhen 518048, China}
\affiliation{Shenzhen Institute for Quantum Science and Engineering, Southern University of Science and Technology, Shenzhen 518055, China}

\author{Shuyi Li}
\affiliation{International Quantum Academy, Shenzhen 518048, China}

\author{Yi-Han Luo}
\affiliation{International Quantum Academy, Shenzhen 518048, China}

\author{Lan Gao}
\affiliation{International Quantum Academy, Shenzhen 518048, China}

\author{Baoqi Shi}
\affiliation{International Quantum Academy, Shenzhen 518048, China}

\author{Chen Shen}
\affiliation{International Quantum Academy, Shenzhen 518048, China}
\affiliation{Qaleido Photonics, Shenzhen 518048, China}

\author{Jijun He}
\affiliation{Key Laboratory of Radar Imaging and Microwave Photonics, Ministry of Education, Nanjing University of Aeronautics and Astronautics, Nanjing 210016, China}

\author{Linze Li}
\affiliation{School of Information Science and Technology, ShanghaiTech University, Shanghai 201210, China}

\author{Tianyu Long}
\affiliation{School of Information Science and Technology, ShanghaiTech University, Shanghai 201210, China}

\author{Baile Chen}
\affiliation{School of Information Science and Technology, ShanghaiTech University, Shanghai 201210, China}

\author{Zhenyu Li}
\affiliation{Institute of Microelectronics, Agency for Science, Technology and Research (A*STAR), Singapore}

\author{Junqiu Liu}
\email[]{liujq@iqasz.cn}
\affiliation{International Quantum Academy, Shenzhen 518048, China}
\affiliation{Hefei National Laboratory, University of Science and Technology of China, Hefei 230088, China}

\maketitle

\section{Characterization of the DFB laser}
\begin{figure*}[h!]
\centering
\includegraphics{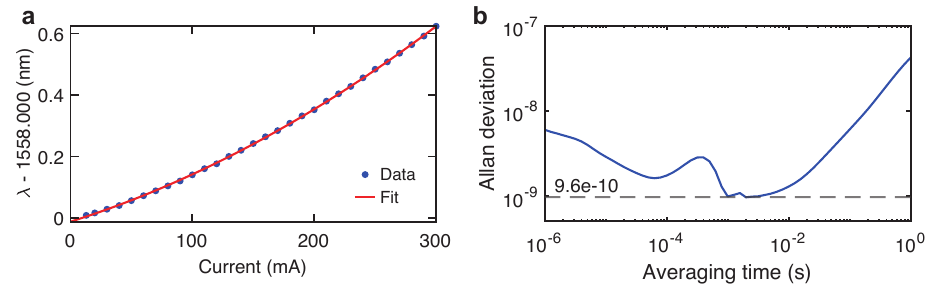}
\caption{\textbf{Characterization of the DFB laser}.
\textbf{a}. 
The DFB laser's emission wavelength versus the driving current.
\textbf{b}. 
Measured Allan deviation data of the DFB laser frequency. 
The fractional frequency instability is $9.6\times 10^{-10}$ at 2 ms.
}
\label{Fig:S1}
\end{figure*}

\vspace{0.5cm}
Supplementary Fig.~\ref{Fig:S1}a shows that, in the 300 mA current range, the DFB laser's emission wavelength can be tuned over 0.6 nm from 1558.0 nm at 25$^\circ$C temperature.
The measured Allan deviation data of the DFB laser's frequency noise are shown in Supplementary Fig.~\ref{Fig:S1}b. 
The free-running DFB laser beats against an external-cavity diode laser (ECDL, Toptica CTL1550). 
The latter's frequency stability is better than the former's. 
The photodetected RF beat signal is captured and analyzed by a phase noise analyzer (PNA), which derives the Allan deviation. 
The measured fractional frequency instability of the free-running DFB laser is $9.6\times 10^{-10}$ at 2 ms. 

\section{Characterization of the silicon Mach-Zehnder modulator}
\vspace{0.5cm}
We use a travelling-wave Mach-Zehnder modulator (TW-MZM) based on Si to convert RF and microwave signals into the optical domain \cite{Li:25}.
The photonic integrated circuit of the Si TW-MZM is a Mach-Zehnder interferometer (MZI), consisting of two multimode interferometers (MMIs) and two waveguide arms, as illustrated in Supplementary Fig.~\ref{Fig:S2}(a, b).
The ground-signal-type electrodes are 2540 $\mu$m long.
The bias voltage $V_\text{dc}$ is applied to measure $V_\pi L_\pi$.
Two push-pull reverse-biased PN junctions are loaded under the electrodes within the waveguide arms.
The PN junctions are formed by doping N and P regions within the arms. 
The electrons and holes in the PN junctions are tuned by electrodes. 
The microwave-to-optic modulation is realized via plasma dispersion effect of electrons' and holes' concentration in Si.
The concentration change results in local change of refractive index ($\Delta n_{\rm Si}$) and optical loss ($\Delta \alpha_{\rm Si}$), described by Soref and Bennett’s equations\cite{Soref:87}, 
\begin{equation}
\begin{cases}
\Delta n_{\rm Si}=-8.8\times10^{-22} n-8.5\times10^{-18}p^{0.8}\\
\Delta \alpha_{\rm Si}=8.5\times10^{-18} n+6.0\times10^{-18} p
\end{cases}
\end{equation}
The $\Delta n_{\rm Si}$ modulates the phase difference between the MZI's two arms, leading to intensity modulation of the optical signal.
This optical modulation is observed as multiple sidebands symmetrically generated around the central signal frequency, detected by an optical spectrum analyzer (OSA).

\begin{figure*}[h!]
\centering
\includegraphics{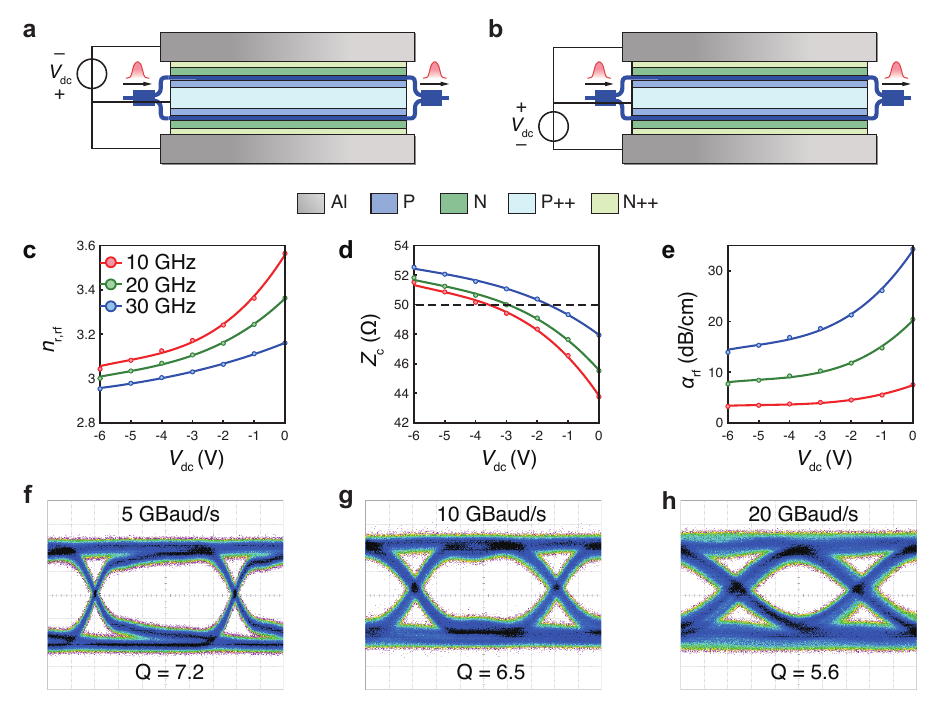}
\caption{
\textbf{Characterization of the Si TW-MZM}.
\textbf{a, b}. 
Schematic when the upper (a) or lower (b) arm of the Si TW-MZM is tuned.
\textbf{c, d, e}.
Simulated phase index $n_\text{r,rf}$, impedance $Z_\text{c}$, and attenuation $\alpha_\text{rf}$ with varying $V_\text{dc}$ and RF frequency $f_\text{rf}=10, 20, 30$ GHz, respectively. 
\textbf{f, g, h}.
Experimental eye diagrams showing measured $Q$ factors of 7.2, 6.5, and 5.6 at data rates of 5, 10, and 20 GBaud/s, respectively.
}
\label{Fig:S2}
\end{figure*}

The design of the Si TW-MZM’s RF electrode adheres to the impedance matching condition to maximize the microwave-to-optic conversion efficiency. 
Supplementary Fig.~\ref{Fig:S2}(c--e) illustrates the electrode’s simulation results, including the phase index $n_\text{r,rf}$, impedance $Z_\text{c}$ and attenuation $\alpha_\text{rf}$, with varying $V_\text{dc}$ and RF frequency $f_\text{rf}=10, 20, 30$ GHz.
The condition $Z_\text{c}=50$ $\Omega$ is satisfied at $V_\text{dc}=-3$ V and $f_\text{rf}=20$ GHz.

To evaluate the microwave-to-optic conversion of the Si TW-MZM, we measure the eye diagrams under non-return-to-zero (NRZ) modulation signals in an optical communication system. 
The quality of these eye diagrams is quantified by the quality factor $Q$ that serves as a measure of the signal-to-noise ratio.
Supplementary Fig.~\ref{Fig:S2}(f--h) presents three clear eye diagrams corresponding to data rates of 5, 10, and 20 GBaud/s, with respective $Q$ factors of 7.2, 6.5, and 5.6. 

\pagebreak
\section{Fabrication and characterization of the Si$_3$N$_4$ microresonator}
\vspace{0.5cm}

The fabrication process flow of Si$_3$N$_4$ integrated waveguides and microresonator is shown in Supplementary Fig. \ref{Fig:S3}a. 
The process is based on 6-inch (150-mm-diameter) wafers using an optimized deep-ultraviolet (DUV) subtractive process \cite{Ye:23, Sun:24}. 
The process starts with deposition of 300-nm-thick Si$_3$N$_4$ on a clean thermal wet SiO$_2$ substrate using low-pressure chemical vapor deposition (LPCVD). 
Afterwards, a SiO$_2$ film is deposited on the Si$_3$N$_4$ layer as an etch hardmask, again using LPCVD. 
After spin-coating of DUV photoresist, DUV stepper lithography based on KrF 248 nm emission is performed to create waveguide pattern on the photoresist mask.
Subsequent dry etching with C$_4$F$_8$, CHF$_3$ and O$_2$ etchants transfers the pattern from the photoresist mask to the SiO$_2$ hardmask, and then to the Si$_3$N$_4$ layer to form waveguides and microresonators. 
The dry etching is optimized to provide smooth and vertical etched sidewall. 
The quality of photolithography and dry etching is critical to minimize optical scattering loss in the waveguides. 

After etching, the photoresist is removed, and the wafer undergoes thermal annealing at 1200 $^\circ$C in nitrogen atmosphere. 
This step is crucial to eliminate hydrogen contents in Si$_3$N$_4$, which cause optical absorption loss. 
Then 3-$\mu$m-thick SiO$_2$ top cladding is deposited on the substrate, followed by another thermal annealing to eliminate hydrogen contents in SiO$_2$. 
Finally, UV photolithography and deep dry etching are performed to define chip size and create smooth chip facets. 
The wafer is separated into individual chips through dicing.

The fabricated Si$_3$N$_4$ microresonators are characterized using a vector spectrum analyzer (VSA) \cite{Luo:24} from 1480 to 1640 nm. 
Supplementary Fig. \ref{Fig:S3}b shows the microresonator's normalized transmission spectrum.  
Each resonance is identified through peak searching and marked with red dots. 
By measuring the frequency of each resonance, the microresonator's integrated dispersion $D_\text{int}$ is obtained.
Each resonance frequency can be expressed as:
\begin{equation}
\begin{aligned}
	\omega_\mu &= \omega_0+D_1\mu+ D_2\mu^2/2+ D_3\mu^3/6+ D_4\mu^4/24\cdots,\\
               &= \omega_0+D_1\mu+D_\text{int}(\mu)
\end{aligned}    
\end{equation}
where $\omega_\mu/2\pi$ is the frequency of $\mu^\text{th}$ resonance relative to the reference resonance ($\mu=0$) of frequency $\omega_0/2\pi$ (the pump laser's frequency),
$D_1/2\pi$ is microresonator FSR, 
$D_2/2\pi$ describes group velocity dispersion (GVD), 
and $D_3$, $D_4$ are higher-order dispersion terms. 
The measured $D_\text{int}$ is shown in Supplementary Fig. \ref{Fig:S3}c, with $D_1/2\pi = 10.699$ GHz and $D_2/2\pi = -88.86$ kHz. 
The negative $D_2$ allows platicon (dark pulse) microcomb generation \cite{Xue:15, Lobanov:15}.

\begin{figure*}[h!]
\centering
\includegraphics{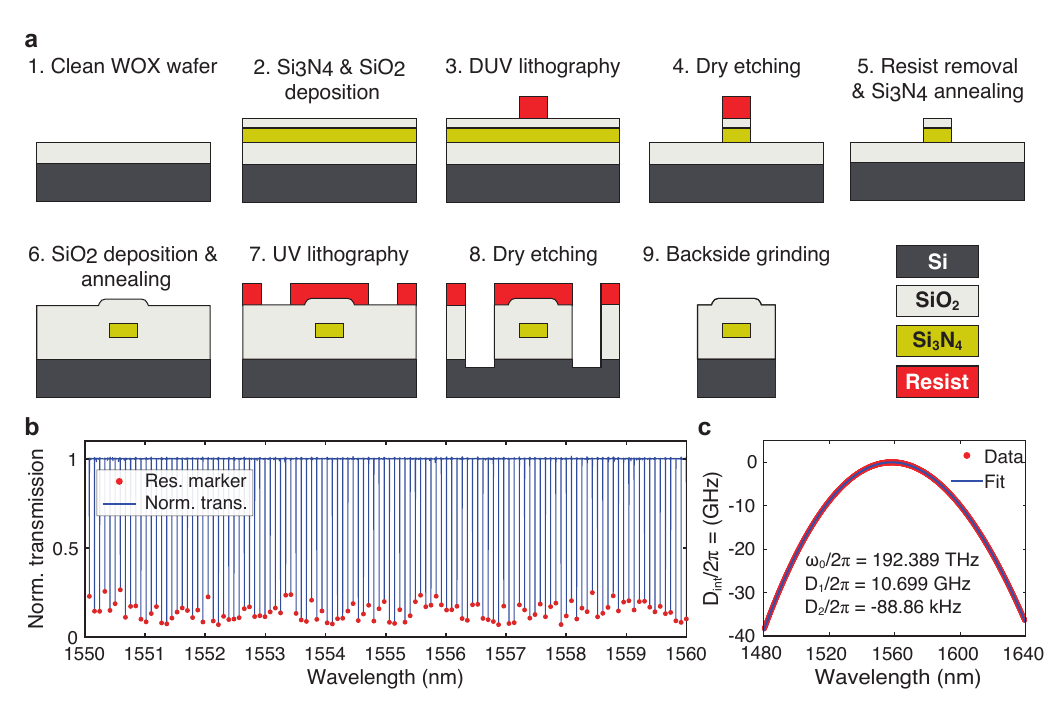}
\caption{
\textbf{Fabrication and characterization of silicon nitride microresonators}.
\textbf{a}. 
The DUV subtractive process flow of 6-inch-wafer Si$_3$N$_4$ foundry fabrication. 
WOX, thermal wet oxide (SiO$_2$).
\textbf{b}.
Measured and normalized microresonator transmission spectrum. 
Identified resonances are marked with red dots.
\textbf{c}.
Measured microresonator's integrated dispersion profile $D_\text{int}$.
The reference frequency is $\omega_0/2\pi=192.389$ THz. 
$D_1/2\pi=10.699$ GHz is the microresonator FSR. 
$D_2/2\pi=-88.86$ kHz is the normal group velocity dispersion. 
}
\label{Fig:S3}
\end{figure*}

\vspace{0.5cm}
\section{Fabrication and characterization of the PD chip}
\vspace{0.5cm}

The high-speed PD chip detects the output optical pulse stream from the Si$_3$N$_4$ microresonator, and generates a microwave signal that is routed to a ground-signal-ground (GSG) probe. 
The epitaxial structure of the PD chip is grown on a semi-insulating InP substrate\cite{LiL:23}, as shown in Supplementary Fig.~\ref{Fig:S4}a. 
The fabrication process starts with P-type contact metals (Ti/Pt/Au/Ti) deposition. 
Dry etching steps are then performed using inductively coupled plasma reactive ion etching (ICPRIE), to form a triple-mesa structure. 
After depositing N-type contact metals (GeAu/Ni/Au), a benzocyclobutene (BCB) layer is coated beneath the coplanar waveguides (CPWs). 
By this approach, the necessity for air-bridge structure can be eliminated, which consequently ensures a consistent and stable connection between p-mesa and CPWs. 

The three-dimensional structure of the PD chip is shown in Supplementary Fig.~\ref{Fig:S4}b.
The active area of the PD chip is 3$\times$15 $\mu m^2$.
The frequency response is measured by an optical heterodyne setup \cite{LiL:23}. 
Supplementary Fig.~\ref{Fig:S4}c shows the voltage-dependent saturation property of the PD measured around 10 GHz.
The ideal relationship between RF power $P_\text{RF}$ and DC photo-current $I_\text{dc}$ is $P_\text{RF} = I_\text{dc}
^2 R_\text{load}/2$, assuming a 100\% 
modulation depth ($I_\text{ac}$ = $I_\text{dc}$), where $R_\text{load}$ = 50 $\Omega$ is the load resistance and $I_\text{ac}$ is AC photo-current.
The measured dark current is below 1 nA, which outperforms commercial high-speed PDs, as shown in Supplementary Fig.~\ref{Fig:S4}d.

\begin{figure*}[h!]
\centering
\includegraphics{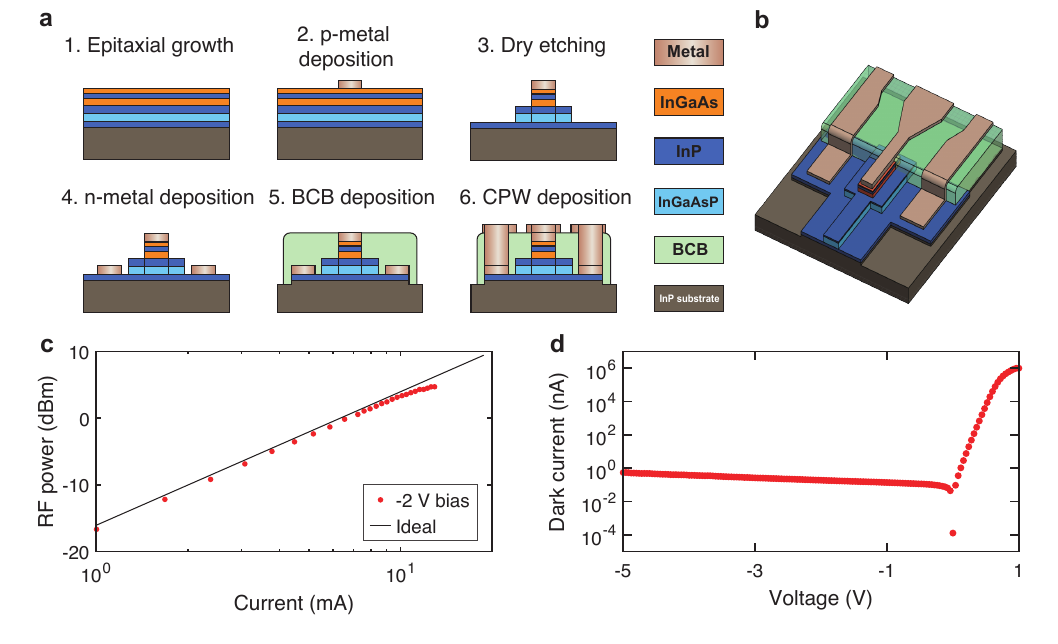}
\caption{
\textbf{Fabrication and characterization of the photodetector chip}.
\textbf{a}. 
Fabrication process flow of the PD chip.
\textbf{b}. 
Three-dimensional structure of the PD chip.
\textbf{c}. 
Measured RF power versus the AC current of PD chip. The data (red dots) are aligned with the ideal case (black line).
\textbf{d}. 
Measured dark current versus bias voltage of the PD chip. 
Negative bias voltage leads to dark current below 1 nA.
}
\label{Fig:S4}
\end{figure*}

\section{Influence of fiber delay length on OEO microwave phase noise}
\vspace{0.5cm}

As illustrated in the main text, the platicon microcomb's $f_\text{rep}$ inherits coherence from the OEO-generated microwave. 
Thus reducing OEO microwave's phase noise improves the coherence of the platicon microcomb. 
Here we study the phase noise property of OEO-generated microwave, without microcomb formation.
Theoretically, the OEO microwave's phase noise can be modelled by \cite{Yao:96}
\begin{equation}
    S_\mathrm{RF}(f) =  \frac{\delta}{(2\pi\tau f)^2},
\label{Equ:PN}
\end{equation}
where $f$ is the Fourier frequency offset relative to the carrier frequency, 
$\delta$ is the noise-to-signal power ratio, 
$\tau$ is the total time delay of the OEO loop (due to the physical path length in the loop and the group delay caused by dispersive elements in the loop). 
Thus, increasing the delay length by 10 times results in nearly 20 dB reduction in the phase noise $S_\mathrm{RF}(f)$. 

\begin{figure*}[h!]
\centering
\includegraphics{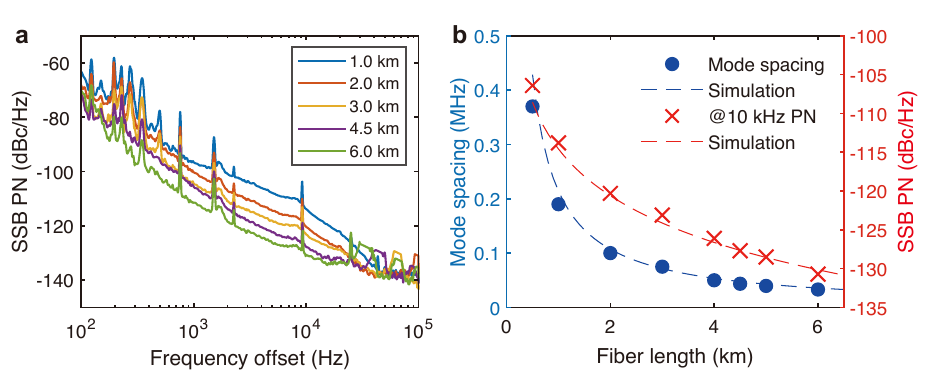}
\caption{
\textbf{OEO microwave with different fiber delay length}.
\textbf{a}. 
Measured OEO microwave's phase noise with different added fiber delay length $L$. 
\textbf{b}. 
For various fiber length, the measured mode spacing (blue dots) in comparison with the analytic trend $\Delta f = c/(n_g L)$ (blue dashed curve), and the measured $S_{L}(10~\text{kHz})$ values (red crosses) in comparison with the analytic trend Eq.~\ref{Equ:PN10} (red dashed curve). 
}
\label{Fig:S5}
\end{figure*}

Supplementary Fig. \ref{Fig:S5}a shows the measured OEO microwave's phase noise with different added fiber delay length, $L=1.0/2.0/3.0/4.5/6.0$ km, respectively.
The measurement is performed by taking a portion of the microwave power from the OEO loop using a power divider. 
The microwave frequency is 10.69927 GHz, and the microwave power injected to the Si MZM is 19 dBm, same as the experiment of OEO platicon. 
We emphasize that, while the OEO loop contains optical fibers and electrical cables, the total delay length is dominated by the total fiber length. 
With this approximation, the OEO microwave's phase noise at 10 kHz Fourier frequency offset $S_{L}(10~\text{kHz})$ can be expressed as
\begin{equation}
    S_{L}(10~\text{kHz}) =  S_{L_0}(10~\text{kHz}) - 20\log_{10}(1+L/L_0),
\label{Equ:PN10}
\end{equation}
where $S_{L_0}(10~\text{kHz})$ is the phase noise with internal delay $L_0$ and without added fiber (i.e. $L=0$). 

The internal delay $L_0$ is mainly caused by the EDFA, and thus is difficult to measure directly. 
Alternatively, $L_0$ can be measured from the OEO oscillation mode spacing $\Delta f = c/(n_g L_0)$, where $n_g$ is the group index of silica single-mode fibers and $c$ is the speed of light in vacuum.
Supplementary Fig. \ref{Fig:S5}b blue dots show the measured mode spacing with various fiber length, in comparison with the analytic trend $\Delta f = c/[(n_g (L+L_0)]$ (blue dashed curve). 
For measured $\Delta f=3.4$ MHz mode spacing, we obtain $L_0=0.06$ km, and the measured $S_{L_0}(10~\text{kHz})=-91$ dBc/Hz.
The measured $S_{L}(10~\text{kHz})$ values with various fiber length $L$ are shown in Supplementary Fig. \ref{Fig:S5}b red crosses, in comparison with the analytic trend Eq.~\ref{Equ:PN10} (red dashed curve). 
The mode spacing $\Delta f$ decreases with increasing fiber length $L$, enhancing modes competition and OEO instability.
Ultimately, we use $L=4.5$ km for OEO platicon experiment as illustrated in the main text.

\section{Purification effect of the platicon on the phase noise}
\vspace{0.5cm}
\begin{figure*}[t!]
\centering
\includegraphics{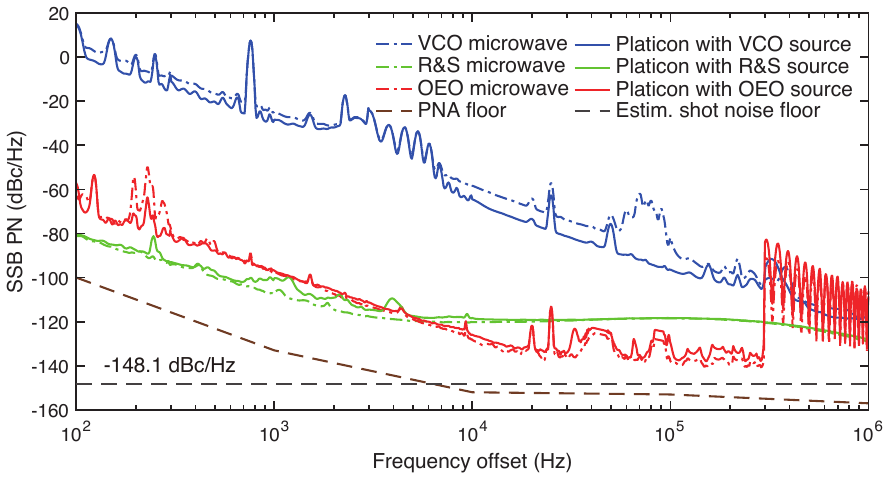}
\caption{
\textbf{Investigation of the platicon purification effect.}
The measured phase noise of the generated platicon's $f_\text{rep}$ is compared to the measured phase noise of the external microwave source that seeds platicon formation.
The dotted blue, green and red curves represent the phase noise data of the 10.7 GHz signals from the VCO, the R\&S microwave generator, and our OEO microwave, respectively. 
The solid blue, green and red curves represent the $f_\text{rep}$ phase noise data of the VCO-modulated platicon,R\&S-modulated platicon, and our OEO platicon, respectively. 
The dashed black curve is the estimated shot noise floor. 
The dashed brown curve is the PNA noise floor.
}
\label{Fig:S6}
\end{figure*}

In Ref.\cite{Liu:22}, it has been observed that the measured phase noise of the generated platicon's $f_\text{rep}$ outperforms the phase noise of the external microwave source which intensity-modulates the CW pump and seeds platicon formation.
This phase-noise purification effect caused by platicon generation is similar to the soliton purification effect observed in Ref. \cite{Weng:19, Liu:20}. 
As the purification effect has not been observed in our OEO platicon, we attribute the reason to the intrinsic ultralow phase noise of our OEO system.
To experimentally verify, we generate platicons via direct intensity-modulation of the CW pump with external microwave sources. 
We compare the measured $f_\text{rep}$ phase noise data of the generated platicons with the phase noise data of the external microwave sources, to quantify the purification effect. 
The microwave sources used here include a common voltage-controlled oscillator (VCO) and a low-noise microwave generator (R\&S SMA100B). 
These measured phase noise data, together with our OEO microwave data, are summarized and compared in Supplementary Fig.~\ref{Fig:S6}.
In the case of the VCO-driving platicon, the platicon purification effect is indeed observed, i.e.  the measured phase noise of the generated platicon's $f_\text{rep}$ outperforms the phase noise of the external microwave source. 
For the platicon driven by the R\&S microwave generator, no phase noise purification is observed, similar to the case of our OEO platicon.

\vspace{0.5cm}
\section{Influence of EA gain on OEO microwave phase noise}
\vspace{0.5cm}

\begin{figure*}[h!]
\centering
\includegraphics{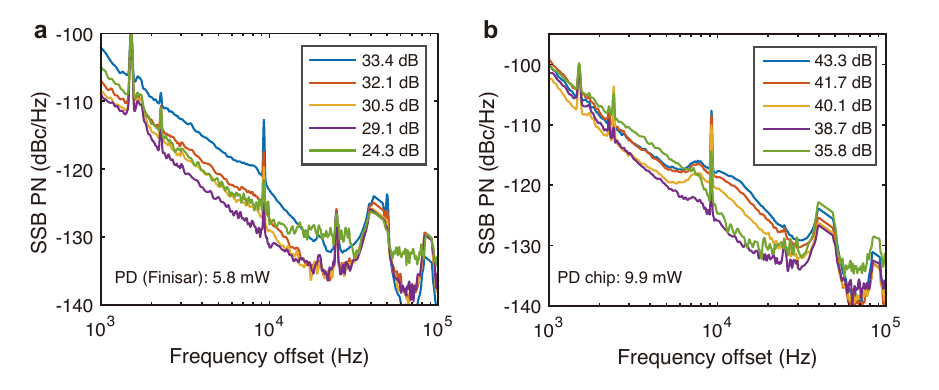}
\caption{\textbf{Influence of EA gain on OEO microwave phase noise}.
Phase noise measurements of OEO microwaves with different EA gain values, using the commercial Finisar PD (\textbf{a}) or the chip PD (\textbf{b}). 
}
\label{Fig:S7}
\end{figure*}

As discussed in Ref. \cite{Chembo:07, Chembo:19}, an inappropriate loop gain increases the OEO microwave's phase noise.
In our experiment, the loop gain is mainly tuned by the gain of the electrical amplifier (EA).
Here we study the influence of EA gain on the OEO microwave's phase noise. 
First, we use a commercial PD (Finisar XPDV3120R-VF-FA) in the OEO loop. 
With 5.8 mW received optical power on the PD, the output microwave power is -8 dBm.
Supplementary Figure \ref{Fig:S7}a shows that, by decreasing the EA gain from 33.4 to 24.3 dB, the lowest phase noise of the OEO microwave is achieved with 29.1 dB EA gain. 
Then we replace the Finisar PD with the chip PD, which generates -20 dBm microwave power with 9.9 mW optical power. 
Supplementary Figure \ref{Fig:S7}b shows that the optimal EA gain is 38.7 dB when using the chip PD. 

\section{Influence of photodetectors, modulators and lasers on the microwave phase noise of OEO platicon}
\vspace{0.5cm}

\begin{figure*}[h!]
\centering
\includegraphics{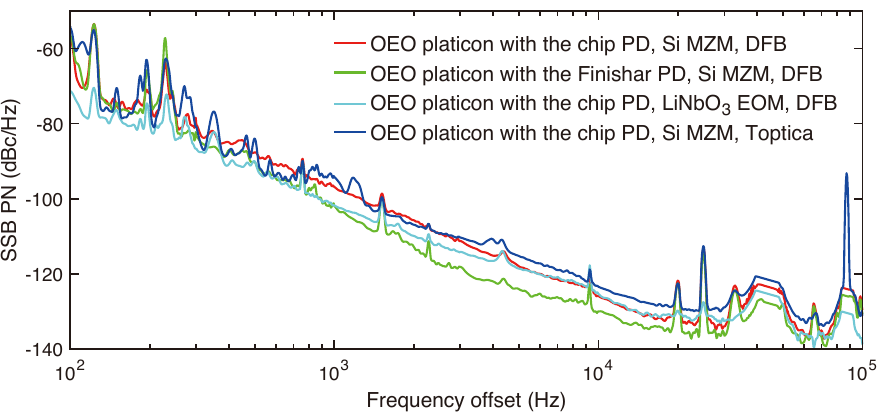}
\caption{
\textbf{Measured microwave phase noise of OEO platicon's $f_\text{rep}$ with different photodetectors, modulators and lasers}.
}
\label{Fig:S8}
\end{figure*}

Here we study the microwave phase noise of OEO platicon's $f_\text{rep}$ with different PDs, modulators and lasers. 
We compare four cases: 
1. OEO platicon with the chip PD, Si MZM, and DFB laser, which is the combination used in the main text; 
2. OEO platicon with the Finishar XPDV3120R-VF-FA PD, Si MZM, and DFB laser; 
3. OEO platicon with the chip PD, LiNbO$_3$ EOM (iXblue MXAN-LN-10-PD), and DFB laser; 
4. OEO platicon with the chip PD, Si MZM, and Toptica CTL laser.
The Si$_3$N$_4$ microresonator is the same in all four cases. 
Supplementary Fig. \ref{Fig:S8} shows the lowest phase noise that can be achieved in all four cases. 
To facilitate quantitative comparison, the components and measured phase noise 
at 0.1/1/10/100 kHz Fourier frequency offset are summarized in Table \ref{tab}. 
We observe phase noise reduction at 0.1 kHz Fourier frequency offset when using the LiNbO$_3$ EOM instead of the Si MZM, and at 10 kHz Fourier frequency offset when using the Finisar PD instead of the chip PD. 
The reasons are likely due to that both the Si MZM and the chip PD are not fully packaged, where optical and microwave power fluctuation on these devices causes loop gain jittering. 

\begin{table*}[h!]
\caption{Comparison highlighting the influence of PDs, modulators and lasers, on the phase noise of OEO platicon's $f_\text{rep}$.}
\begin{ruledtabular}
\begin{tabular}{l@{}ccccccc}
\multirow{2}*{Color} & \multirow{2}*{Photodetector} & \multirow{2}*{Modulator} & \multirow{2}*{Laser} & dBc/Hz @ & dBc/Hz @ & dBc/Hz @ & dBc/Hz @ \\
 & & & & 0.1 kHz & 1 kHz & 10 kHz & 100 kHz \\
\colrule\\[-6pt]
Red & chip PD & Si MZM & DFB & $-57$ & $-97$ & $-126$ & $-130$\\
Green & Finisar PD & Si MZM & DFB & $-57$ & $-102$ & $-130$ & $-131$\\
Cyan & chip PD & LiNbO$_3$ EOM & DFB & $-71$ & $-102$ & $-125$ & $-139$\\
Blue & chip PD & Si MZM & Toptica & $-55$ & $-94$ & $-123$ & $-131$\\
\end{tabular}
\end{ruledtabular}
\label{tab}
\end{table*}

\newpage
\vspace{1cm}
\section*{Supplementary References}